\documentclass[aip, amsmath, amssymb, reprint]{revtex4-2}

\usepackage{graphicx}
\usepackage{dcolumn}
\usepackage{hyperref}
\usepackage{bm}

\usepackage[utf8]{inputenc}
\usepackage[T1]{fontenc}
\usepackage{mathptmx}
\usepackage{gensymb}

\usepackage{ulem} 

\hypersetup{
    colorlinks=true,
    linkcolor=blue,
    citecolor=blue,
    filecolor=blue,
    urlcolor=blue
}

\graphicspath{{./}{./figs/}}

\begin{document}


\title[]{Comparison of two multiplexed portable cold-atom vacuum standards}

\author{Lucas H. Ehinger}
\affiliation{Department of Physics, 901 12$^{\rm th}$ Avenue, Seattle University, Seattle, Washington 98122, USA}
\author{Bishnu P. Acharya}
\affiliation{
Sensor Science Division, National Institute of Standards and Technology, Gaithersburg, Maryland 20899, USA
}
\author{Daniel S. Barker}
\affiliation{
Sensor Science Division, National Institute of Standards and Technology, Gaithersburg, Maryland 20899, USA
}
\author{James A. Fedchak}
\affiliation{
Sensor Science Division, National Institute of Standards and Technology, Gaithersburg, Maryland 20899, USA
}
\author{Julia Scherschligt}
\affiliation{
Sensor Science Division, National Institute of Standards and Technology, Gaithersburg, Maryland 20899, USA
}
\author{Eite Tiesinga}
\affiliation{
Joint Quantum Institute, College Park, Maryland 20742, USA
 and National Institute of Standards and Technology, Gaithersburg, Maryland 20899, USA
}
\author{Stephen Eckel}
\email{stephen.eckel@nist.gov}
\affiliation{
Sensor Science Division, National Institute of Standards and Technology, Gaithersburg, Maryland 20899, USA
}

\date{\today}

\begin{abstract}
We compare the vacuum measured by two portable cold-atom vacuum standards (pCAVS) based on ultracold $^7$Li atoms.
The pCAVS are quantum-based standards that use {\it a priori} scattering calculations to convert a measured loss rate of cold atoms from a conservative trap into a background gas pressure.
Our pCAVS devices share the same laser system and measure the vacuum concurrently.
The two pCAVS together detected a leak with a rate on the order of $10^{-6}$~Pa~L/s.
After fixing the leak, the pCAVS measured a pressure of about 40~nPa with 2.6~\% uncertainty.
The two pCAVS agree within their uncertainties, even when swapping some of their component parts.
Operation of the pCAVS was found to cause some additional outgassing, on the order of $10^{-8}$~Pa~L/s, which can be mitigated in the future by better thermal management.
\end{abstract}


\maketitle

\section{Introduction}
Cold-atom vacuum standards promise to deliver accurate measurements of vacuum in the ultra-high vacuum (UHV, $<10^{-6}$~Pa) and extreme-high vacuum (XHV, $<10^{-9}$~Pa) regimes.\cite{Makhalov2016, Scherschligt2017, Eckel2018, Booth2019, Shen2020}
In these regimes, hot-cathode ionization gauges, such as Bayard-Alpert gauges and their derivatives, are the typical means of pressure measurement.\cite{Bayard1950,Akimichi1996,Redhead1966}
Yet they suffer from several known systematics, including unwanted X-ray-induced currents, electron stimulated desorption of ions and neutrals, and thermal outgassing.
Cold-atom-based vacuum gauges, on the other hand, should alleviate most of these systematics, while also being a primary standard.
Thus far, all cold-atom vacuum standards have been laboratory-scale devices and are not suitable replacements for an ionization gauges.\cite{Makhalov2016, Booth2019, Shen2020}

Here, we directly compare two portable cold-atom vacuum standards (pCAVS) based on ultracold lithium, the main components of which have been used to demonstrate a compact apparatus for laser cooling and trapping of strontium.\cite{Sitaram2020}
The gauge head (not including the laser system) fits within a roughly 15~cm $\times$ 35~cm $\times$ 50~cm cuboid; the vacuum components of our standard comprise a total volume of approximately $1.3$~L.
The cost of the cold-atom vacuum standard is primarily driven by the laser system used to cool, trap, and count the atoms.
The laser system is shared between both pCAVSs through an optical fiber switch.
Using the hardware tested here, up to four pCAVS devices can be multiplexed off a single 450~mW laser.
We estimate the uncertainty of our portable gauges.
The reproducibility of the pCAVS is estimated, in part, by swapping components between the two.
Finally, we demonstrate that by comparing two pCAVS, we can detect small vacuum leaks with rates of the order of $10^{-6}$~Pa~L/s.
Given the sensitivity of the pCAVS, we estimate that we can detect leaks with rates as small as $10^{-9}$~Pa~L/s.

Cold-atom vacuum gauges, in their simplest form, measure the loss rate $\Gamma$ of ultracold ($\lesssim 500$~$\mu$K) sensor atoms from a trap induced by collisions with background gas atoms or molecules (typically near 300~K).
If all other sources of sensor atom loss are small (including Majorana losses, three-body recombination losses, collisional spin-flip losses,  etc.) and the cold-atom trap depth is negligible, then $\Gamma = K n$, where $K$ is the total elastic and inelastic rate coefficient and $n$ is the number density of background gas atoms in the vacuum chamber.
Inserting the ideal gas law, valid in the UHV and XHV, the pressure is then related to the measured decay rate through
\begin{equation}
    \label{eq:ideal_meas_equation}
    p =  n kT= \frac{\Gamma}{K}kT\,,
\end{equation}
where $k$ is the Boltzmann constant and $T$ is the temperature of the background gas.

The rate coefficient $K$ is species dependent and typically of the order of $10^{-9}$~cm$^3$/s.
It can be estimated for different gases using semiclassical collision theory first developed in the 1960s.\cite{Child,Eckel2018}
Accurate values from {\it a priori} quantum collision calculations are more sparse.
For 300~K H$_2$, the dominant background gas in the UHV and XHV domains, colliding with ultracold $^7$Li sensor atoms, this rate coefficient has been calculated {\it a priori} and is $K_{\rm H_2}=3.18(6)\times10^{-9}$~cm$^3$/s.\cite{Makrides2019,Makrides2022Eb}
At 300~K only four rotational levels of the ground vibrational level of H$_2$ are significantly populated.
Similar calculations have also been completed for 300~K He colliding with ultracold $^7$Li showing $K_{\rm He}=1.659(15)\times10^{-9}$~cm$^3$/s.\cite{Makrides2020,Makrides2022Ea}
Because of this underpinning theory, cold-atom vacuum gauges are primary standards, being traceable to the SI second and kelvin.\cite{Scherschligt2017}

\begin{figure}
    \centering
    \includegraphics[width=\columnwidth]{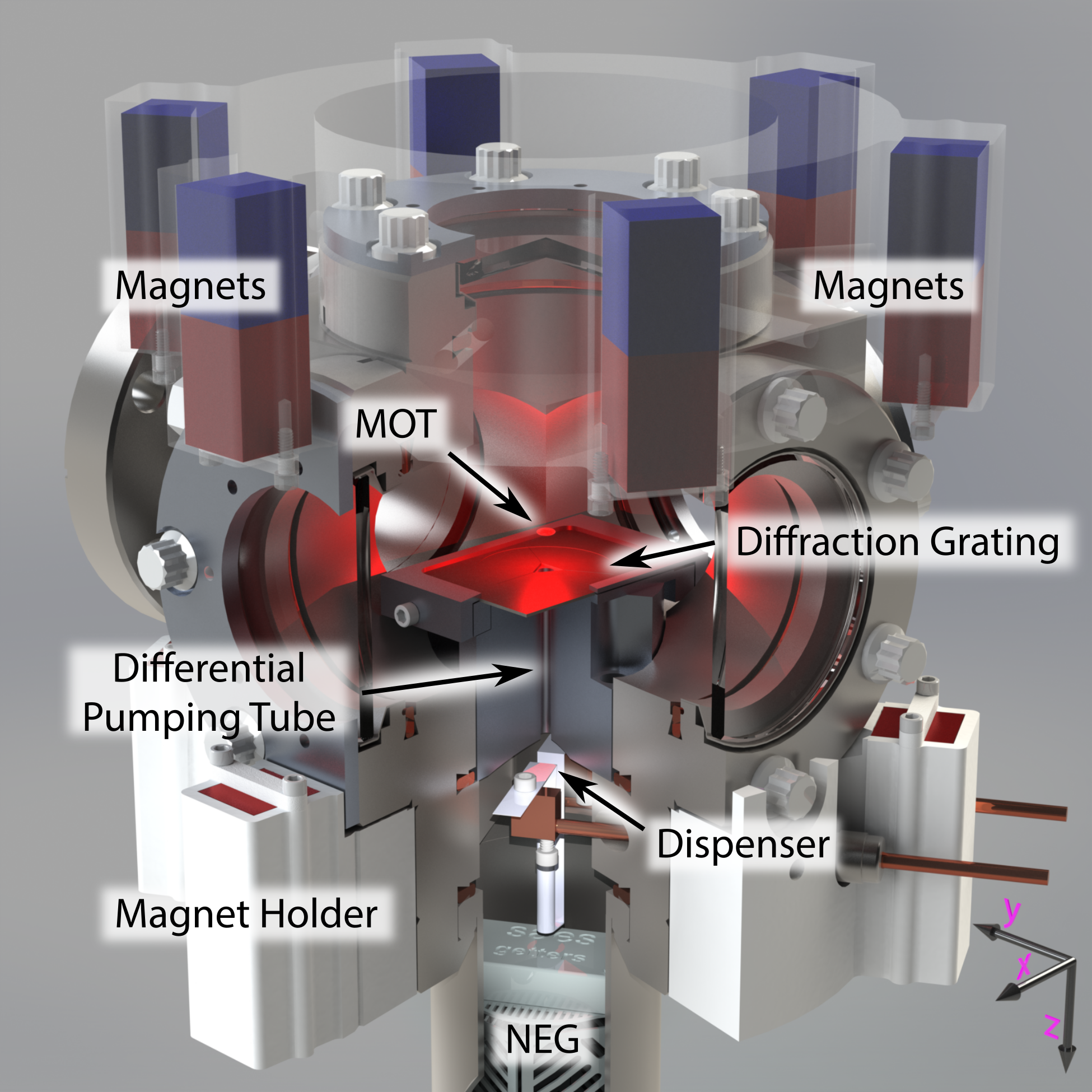}
    \caption{A computer-assisted-design rendering of one of our portable cold-atom vacuum standards (pCAVS).  For scale, the height of one of the magnets is 3.81~cm.  Not shown is the titanium shutter, which fits snugly between the dispenser and the differential pumping tube. A non-evaporable getter (NEG) pumps the source chamber.  The red and blue coloring of the magnets indicate the relative orientation of their poles.}
    \label{fig:sketch}
\end{figure}

\section{Description}\label{sec:description}

The pCAVS is designed to reduce the size of the vacuum components, simplify optical alignment, and minimize outgassing into the vacuum being measured by the source of atoms.
Figure~\ref{fig:sketch} shows a computer-assisted-design rendering of a pCAVS.
It is divided into two sections: the source chamber that produces a beam of thermal atoms and a measurement chamber where a magneto-optical trap (MOT) is formed and the vacuum pressure is measured.
The two sections are separated by a differential pumping tube of length 2.67~cm and radius 1.5~mm, yielding a conductance of approximately $0.014$~L/s for N$_2$ at 300~K.
The conductance for other gases is similar, as it scales as $\sqrt{m}$, where $m$ is the mass of the background gas atom or molecule.

The source chamber contains a commercial, stainless steel alkali-metal dispenser (AMD) that heats and vaporizes lithium atoms, causing them to effuse from the source toward the differential pumping tube.
The AMD is heated by running a direct current $i$ that dissipates between 2~W and 6~W in the AMD.
For most data in this article, the two AMDs in the two pCAVS are connected in series, and thus have the same $i$.
A 3D-printed titanium shutter, actuated by a small, magnetic rotary feedthrough driven by a stepper motor, can stop effusing atoms by completely blocking the differential pumping tube.
In the closed position, the shutter does not form a vacuum-tight seal, but is only 0.25~mm from the end of the differential pumping tube.
Such a small distance further reduces the effective conductance of the differential pumping tube.
Because the warm AMD will emit species in addition to lithium through thermal outgassing into the vacuum, a non-evaporable getter (NEG) pump with 100~L/s pumping speed for H$_2$, the dominant background gas, is installed just below the AMD. 
Pumping speeds for other chemically reactive gas species are smaller; non-reactive noble gases must be pumped through the differential pumping tube.

The measurement section is built around a nanofabricated grating MOT chip,\cite{Nshii2013, Barker2019, Sitaram2020} designed and fabricated at NIST.
The input beam for the grating is produced using bulk optics.
Each pCAVS uses one of two different input-beam shaping systems that produce a collimated, circularly-polarized beam with approximately 20~mm $1/e^2$ diameter.
The first has a total length of 24~cm and collimates light that expands from a polarization maintaining fiber with a 150~mm lens.
The second has a slightly shorter total length of 20~cm and uses a $-9$~mm focal length lens to force the beam to diverge more quickly from the fiber, then uses a secondary lens with focal length 100~mm to collimate the beam.\cite{Imhof2017}
Future designs can shorten the input-beam-shaping systems by either incorporating more lenses or switching to planar optics.\cite{McGehee2021}
The laser beam from each input-beam-shaping system is aligned such it is normally-incident to and centered on the grating chip.
We note that though the laser beam is well-aligned to the chip, the chip normal and the differential pumping tube axis may be out of parallel by several degrees.
Any misalignment reduces the efficacy of laser slowing as lithium atoms pass through the differential pumping tube, consequently causing a reduction of the MOT loading rate.\cite{Barker2019}

Type-N52 NdFeB permanent magnets mounted to the exterior of the chamber with 3D-printed holders generate a spherical quadrupole magnetic field used by both the MOT and subsequent magnetic trap.
The zero field position is $z_{\rm T}=6.2(1.0)$~mm above the diffraction grating chip, where the standard uncertainty in parenthesis comes from an estimated 2~mm positioning uncertainity in the vertical direction.\footnote{Unless stated otherwise, all uncertainties quoted in this paper are standard ($k=1$) uncertainties.}
(We cannot directly measure the location of the zero of the magnetic field, as it does not coincide with the center of the MOT, and do not have {\it in situ} absorption imaging available to image the atoms while trapped in the magnetic trap.\cite{McGilligan2015})
Figure~\ref{fig:sketch} shows a configuration of magnets and their respective holders that generates an axial gradient of $B' =4.59(17)$~mT/cm, normal to the chip.
A second configuration, shown in Ref.~\onlinecite{Sitaram2020}, containing more magnets and using different holders generates a larger axial gradient of  $B'=7.53(28)$~mT/cm.

The laser light is about $-9$~MHz  detuned from the  $^{2}$S$_{1/2}(F=2)\rightarrow ^{2}$P$_{3/2}(F'=3)$ main cycling transition in $^7$Li, which has a natural linewidth $2\pi\times5.87$~MHz.
Due to the unresolved hyperfine structure of the excited $^2$P$_{3/2}$ state, atoms can fall out of the cycling transition and into the $^{2}$S$_{1/2}(F=1)$ state.
To pump atoms out  of this ${F=1}$ state, an electro-optic modulator (EOM) is inserted into the laser beam and produces sidebands at $\pm 813$~MHz. The blue sideband is resonant to the $^2$S$_{1/2}(F=1)\rightarrow ^2$P$_{3/2}(F'=2)$ transition and has roughly 50~\% of the power of the carrier, as measured with a Fabry-Perot cavity.
The laser light is then inserted into a 1$\times$4 fiber switch that is used to select to which pCAVS the  light is sent.
The switch operates with $6$~dB of loss and requires approximately 100~ms to actuate between the output ports of the switch. 
The switch outputs are sent to the two beam-shaping systems, where we measure approximately 20~mW of total power and a peak intensity of 12~mW/cm$^2$, approximately four times the saturation intensity of the main cycling transition.

The two pCAVSs are mounted to the same vacuum chamber, which is pumped by an ionization pump with an N$_2$ pumping speed of 60~L/s.
We estimate the conductance between the pCAVSs and the ionization pump to be roughly 10~L/s for N$_2$.
The atom clouds in the two pCAVSs are separated by 20~cm and have direct line of site to each other.
The nominal vacuum pressures at the two atom clouds should be identical.

To obtain the lowest pressures possible, we applied two treatments to the vacuum components.
Prior to assembly, all stainless steel components were baked in a vacuum furnace at 400~$^\circ$C for roughly 40~d to eliminate as much dissolved hydrogen as possible.~\cite{Sefa2017, Fedchak2018a}
After assembly, the evacuated system was baked at 150~$^\circ$C for a few days to desorb water from surfaces and then allowed to cool down to ambient temperature $T$.

We measure the background gas temperature $T$ using two calibrated industrial-quality platinum-resistance thermometers (PRTs).
We assume the background gas is in thermal equilibrium with the chamber walls.
One PRT is mounted directly onto the exterior pCAVS \#1 roughly centered vertically in the measurement section; the other is mounted to the exterior of the vacuum chamber between the two pCAVS.
Self-heating effects are negligible at our level of uncertainty of 1~K to 2~K.

\section{Measurements}

The following procedure measures the pressure in either pCAVS.
First, the shutter is opened, the EOM turned on, and laser light is injected into the pCAVS, forming the MOT.
The MOT is loaded for about 2~s, at which time it contains a modest $10^5$ atoms.
After the MOT is loaded, the shutter is closed.
The EOM is then turned off, removing light resonant with the $F=1\rightarrow F'=2$ transition, and all $^7$Li atoms transfer into the $^2$S$_{1/2}(F=1)$ hyperfine state.
The laser is turned off, and only atoms with projection quantum number $m_F=-1$ with respect to the local magnetic field direction will congregate where the magnetic field strength is zero.
Typically, of the order of $10^4$ atoms are initially trapped in this magnetic quadrupole trap.
The atoms are held in the quadrupole trap for a variable time $t$, after which the laser and EOM are turned on for 15~ms, reforming the MOT and capturing the remaining atoms from the quadrupole trap.
The MOT is then imaged for 20~ms to count the number of atoms, $N$.
After this procedure of measuring $N$ at a single $t$ is completed on the first pCAVS, the laser is switched to address the second pCAVS.
To measure a full decay curve, the full procedure with both pCAVS is repeated several times at different $t$.
Thus, the pressure in each pCAVS is measured in an interlaced pattern.

\begin{figure}
    \includegraphics[trim=0 0 0 0,clip]{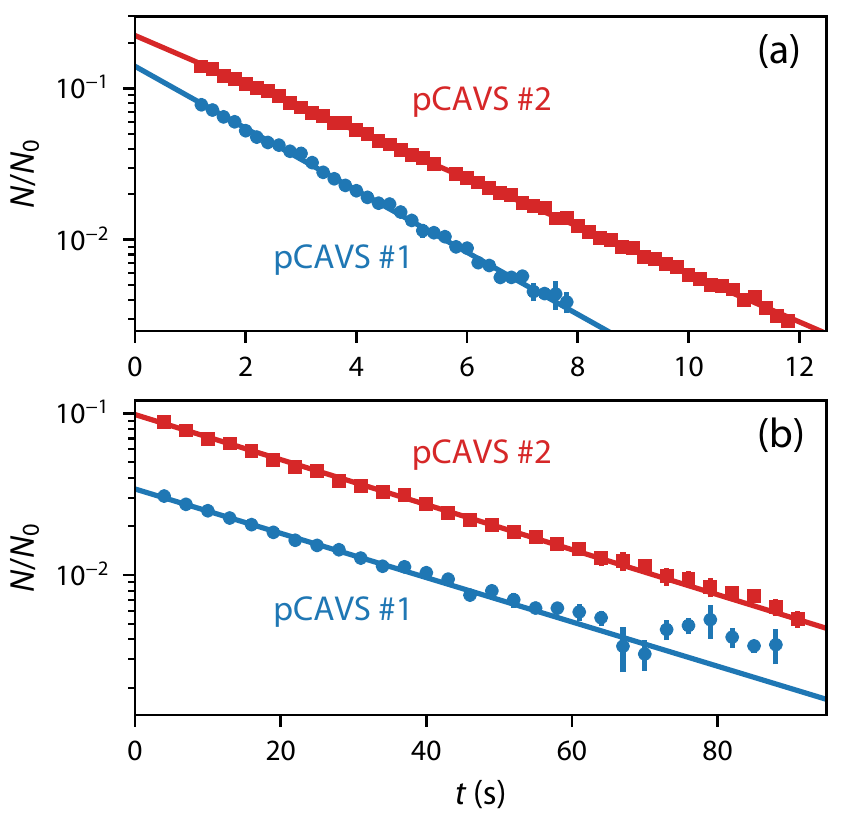}
    \caption{Recaptured $^7$Li atom number $N$, normalized to the separately measured just after the MOT loading phase $N_0$, as functions of time $t$. Blue circles are for pCAVS \#1; red squares represent pCAVS \#2. Solid lines show fits assuming exponential decay. These example decays were recorded before (a) and after (b) a leak in pCAVS \#1 was fixed; note the difference in time scales.  The error bars on the markers correspond to the standard error in the mean.}
    \label{fig:example_decays}
\end{figure}

Our initial measurements with the two pCAVSs are shown in Fig.~\ref{fig:example_decays}(a).
The graph shows the number of recaptured atoms $N$ normalized to the number of atoms, $N_0$, observed in the MOT just after the loading phase, as a function of $t$.
Shot-to-shot fluctuations in $N/N_0$ are up to 20\,\% smaller than those in $N$.
These data represent the mean of four repeated measurements and were taken at an AMD current of $i=5.9$~A. The error bars in the figure are standard uncertainties in the mean of the repeated measurements (type A, statistical uncertainty).
As can be  seen from Fig.~\ref{fig:example_decays}(a), the two pCAVS have different atom-number decay times and thus, measure different pressures.
Each of the curves is fit to $N/N_0= A e^{-\Gamma t}$, where $A$ and $\Gamma$ are adjusted parameters.
Least-square fits assuming uncorrelated  uncertainties at
different $t$ yield $\Gamma = 0.471(4)$~s$^{-1}$ with reduced $\chi_\nu^2$ equal to $1.0$ and
$\Gamma =0.362(2)$~s$^{-1}$ with reduced $\chi_\nu^2$ equal to $10$ for pCAVS \#1 and pCAVS \#2,
respectively. The number of degrees of freedom $\nu$ are $32$ and $51$, respectively.
The large reduced $\chi_\nu^2$ for pCAVS \#2 is caused by an underestimate of the uncertainty of $N/N_0$ at five times $t$, presumably due to using only four repeats to determine the standard error in the mean.
We account for this implied systematic uncertainty by multiplying the covariance matrix for parameters $A$ and $\Gamma$ by reduced $\chi_\nu^2$, {\it i.e.}, increasing the uncertainty of $\Gamma$ for pCAVS \#2 to the quoted 0.002~s$^{-1}$.

We examine potential causes for the different decay rates including: the AMDs, the input beam-shaping systems, and the magnets and their respective holders.
The AMDs could dissipate different amounts of heat, causing them to have different temperatures and outgassing rates.
Different outgassing rates could give rise to a pressure gradient between the two pCAVS devices.
We tested this by measuring the decay rate as a function of current $i$, but did not observe a statistically significant change between 6~A to 8~A.
(Detailed measurements of pressure as a function of $i$ will be presented later.)
We then swapped the input beam-shaping systems.
At constant current, the decay rates measured by the two pCAVSs were statistically unchanged with the ratio of measured $\Gamma$s before and after exchanging input beam-shaping systems equal to 0.90(5) and 0.93(8) for pCAVS \#1 and pCAVS \#2, respectively.
Finally, we increased the magnetic field gradient from 4.6~mT/cm to 7.5~mT/cm, and the ratio of the measured $\Gamma$s was 1.04(5) and 1.00(8), for pCAVS \#1 and pCAVS \#2, respectively.

One must also consider that the two pCAVS were {\it simply} at different pressures.
A leak could cause such a discrepancy, and we can estimate the leak rate using our two pCAVS devices.
Because the leak gas is mostly N$_2$ (air) and using the semi-classical estimate of $K_{{\rm N}_2}=2.5\times10^{-9}$~cm$^3$/s at a temperature $T=295$~K,\cite{Eckel2018}
the N$_2$ pressure is $7\times10^{-7}$~Pa and $6\times10^{-7}$~Pa
in  pCAVS \#1 and \#2, respectively, giving a pressure difference of $1\times10^{-7}$~Pa.
With the 10~L/s estimated pumping speed on each pCAVS, the leak rate is then of the order of $10^{-6}$~Pa~L/s (equivalent to a flow of the order of  $10^{-13}$~mol/s), small enough to evade detection with a residual gas analyzer using a Faraday cup detector.
Attaching a Hiden HAL RC 201 residual gas analyzer with an electron multiplier to the vacuum system, we found the leak via helium leak testing and verified the estimate of the leak rate.\footnote{Certain commercial equipment, instruments, or materials are identified for reference purposes only.  Such identification is not intended to imply recommendation or endorsement by NIST, nor is it intended to imply that the materials or equipment identified are necessarily the best available for the purpose.}

After repairing the leak, the two pCAVS measure equal decay times, as shown in Fig.~\ref{fig:example_decays}(b), for $i=7.5$~A and $T=301.7(1.6)$~K.
Once again, each data point is an average of $N/N_0$ of four repeats; the error bar is the $k=1$ standard uncertainty in the mean.
Least-square fits to an exponential as function of $t$ assuming uncorrelated uncertainties yield $\Gamma = 0.0316(6)$~s$^{-1}$ with reduced $\chi_\nu^2$ equal to $3.8$ and $\Gamma =0.0321(3)$~s$^{-1}$ with reduced $\chi_\nu^2$ equal to $0.7$ for pCAVS \#1 and pCAVS \#2, respectively.
The number of degrees of freedom $\nu$ are $29$ and $28$, respectively.
Here, the large reduced $\chi_\nu^2$ for pCAVS \#1 comes from the large scatter of points at times $t>60$~s, where $N$ becomes close to our atom-number detection threshold.
As before, we  account for this implied systematic uncertainty by increasing the covariance matrix for parameters $A$ and $\Gamma$ by reduced $\chi_\nu^2$, {\it i.e.}, increasing the uncertainty of $\Gamma$ for pCAVS \#1 to the quoted 0.0006~s$^{-1}$.
The two pCAVS agree at the $k=1$ uncertainty level.

Let us now calculate the pressure in the vacuum chamber and its uncertainty including potential systematic shifts inherent in the operation of the pCAVSs.
We can safely assume that the background gas is H$_2$, as it is typically the dominant contaminant of a vacuum chamber in the UHV and XHV regimes when no leaks are present.
We modify Eq.~\ref{eq:ideal_meas_equation} to include ``glancing'' collisions, i.e, collisions that do not transfer enough energy to eject an atom from the trap, and other loss mechanisms inherent to the quadrupole trap and resulting from collisions between cold $^7$Li atoms.
With these modifications, the pressure $p$ is found through
\begin{equation}
    \label{eq:meas_equation}
    p = \frac{\Gamma - \Gamma_{\rm other}}{K_{{\rm H}_2}\times [1-f_{\rm gl}(U)]} k T,
\end{equation}
where $\Gamma_{\rm other}$ captures other loss mechanisms and $f_{\rm gl}(U)$ is the fraction of glancing collisions given a quadrupole trap depth $U$.
At our level of uncertainty, we can safely neglect both the background gas temperature and $^7$Li temperature dependence of the rate coefficient $K_{\rm H_2}$ given in the introduction.

For the data in Fig.~\ref{fig:example_decays}(b), the measured temperature of the vacuum chamber is 301.7(1.6)~K.
The standard uncertainty $u(T)$ is a combination of an observed $1.3$~K gradient between our two platinum-resistance thermometers, a $0.8$~K common drift over the full measurement duration ($\approx 3.5$~h), and a small $0.03$~K  calibration uncertainty.
The uncertainty in $\Gamma$ is statistical (type-A), and its relative value $u(\Gamma)/\Gamma=0.022$  and 0.01 for pCAVS \#1 and pCAVS \#2, respectively.
(Any timing uncertainty is negligible at this level of precision.)

Unlike for $K_{{\rm H}_2}$, no fully quantum mechanical scattering calculation exists for the fraction of glancing collisions $f_{\rm gl}(U)$.
However, $f_{\rm gl}(U)$ can be estimated from a semiclassical theory that has as its only molecular input the van-der-Waals coefficient $C_6$ of the attractive long-range dispersion potential between Li and H$_2$.\cite{Child}
Restating the results of Ref.~\onlinecite{Booth2019} in terms of length and energy scales for a van-der-Waals potential, we find
\begin{equation}
    \label{eq:universality2}
    f_{\rm gl}(U) = 0.179\cdots \times
     \left(\frac{kT_{\rm eff}}{E_6}\right)^{-1/5} \frac{m_{\rm Li}}{\mu}\frac{U}{E_6} + O(U^2)\,,
    \end{equation}
where length $x_6=\sqrt[4]{2\mu C_6/\hbar^2}$, energy $E_6=\hbar^2/2\mu x_6^2 $,
$\mu=m_{\rm Li} m_{{\rm H}_2}/(m_{\rm Li} + m_{{\rm H}_2})$ is the reduced mass of the $^7$Li+H$_2$ system, and $\hbar$ is the reduced Planck constant.
Here, $m_{i}$ with $i={\rm Li}$ and H$_2$ are the masses of  $^7$Li and H$_2$, respectively.
The  temperature $T_{\rm eff}= m_{\rm Li} T/(m_{\rm Li}+m_{{\rm H}_2})<T$ is the effective temperature in the relative motion between $^7$Li and H$_2$ accounting for the very different temperatures of $^7$Li and H$_2$.
We have $E_6/k = 80.5$ mK for $^7$Li+H$_2$
and estimate that the semiclassical theory is accurate to within 50~\%.

The trap depth $U$ of the quadrupole  trap for atoms in the $m_F=-1$ projection state is given by $U =- g_F  \mu_B B' z_T$, where $g$ factor $g_F=-0.50$ and $z_T=6.2(1.0)$~mm is the nearest distance to the chip.\cite{Eckel2018}
For our pCAVS systems, $U/k=1.5(3)$~mK, making $f_{\rm gl} = 3.2(1.7) \times 10^{-3}$.
Here, the uncertainty is the uncorrelated combination of the uncertainty of $U$ and our estimate of the accuracy of the semiclassical theory.
Only the term shown in Eq.~(\ref{eq:universality2}), the expansion of $f_{\rm gl}(U)$ in $U$ around $U=0$, is required at our level of precision.

Several sensor atom loss mechanisms contribute to $\Gamma_{\rm other}$.
Majorana losses (losses due to non-adiabatic $m_F$ spin flips near the magnetic field zero) have a characteristic timescale of $\Gamma_{\rm Majorana}\sim \hbar/m_{\rm Li} r_c^2$, where $r_c$ is the characteristic size of the cloud.\cite{Petrich1995, Eckel2018}
Here, we take $r_c$ to be the $1/e$ radius of the cloud along the radial direction.
For $B$-field gradient $B'=7.53$~mT/cm and an estimated $^7$Li temperature of $750$~$\mu$K,\cite{Barker2019} $r_c\approx6$~mm and we derive that $\Gamma_{\rm Majorana} \approx 3.6\times10^{-4}$~s$^{-1}$,  1~\% of the measured $\Gamma$.
The functional form for Majorana loss is unknown (it may be super-exponential) and further study is required.
We conservatively assume $\Gamma_{\rm Majorana}=0$ and $u(\Gamma_{\rm Majorana}) = 3.6\times10^{-4}$~s$^{-1}$.

The next largest contribution to $\Gamma_{\rm other}$ is thermalization or evaporative loss (due to elastic collisions between two cold $^7$Li atoms that cause one atom to be lost from the quadrupole trap).
With $10^4$ atoms in the quadrupole trap at $750$~$\mu$K, our  peak density is of the order of $10^5$~cm$^{-3}$ and evaporative loss rates are $\lesssim 10^{-6}$~s$^{-1}$.\cite{Luiten1996}
Collisions between cold $^7$Li atoms that induce spin flips or three-body recombination and cause ejection are further suppressed, occurring with rates $\lesssim 10^{-11}$~s$^{-1}$.
Hence, we take $\Gamma_{\rm other}=0$ and $u(\Gamma_{\rm other})=u(\Gamma_{\rm Majorana})$.

\begin{table}
    \centering
    \begin{tabular}{|c|cc|}
    \hline
          & Pressure & Pressure \\
         Contribution/Source & Independent & Dependent \\
         \hline
         Rate coefficient, $K_{{\rm H_2}}$ & & $0.019p$\\
         Temperature, $T$ & & $0.005p$ \\
         Glancing fraction, $f_{\rm gl}$ & & $1.7\times10^{-3} p$ \\
         Majorana losses, $\Gamma_{\rm other}$ & 0.33~nPa & \\
         Statistical, $\Gamma$ & 0.39~nPa & \\
         \hline
         Total Variance & \multicolumn{2}{c|}{$(0.51~\mbox{ nPa})^2 + (0.020p)^2$} \\
         \hline
    \end{tabular}
    \caption{One-standard-deviation, $k=1$ uncertainty budget and total variance of the pressure measured by pCAVS \#2 using $^7$Li sensor atoms for the decay shown in Fig.~\ref{fig:example_decays}(b).  Quantity $p$ in the right-most column is the measured pressure in the vacuum chamber; all other symbols correspond to those of Eq.~\ref{eq:meas_equation}. For the total variance  the uncertainties are added in quadrature assuming no correlations among the contributions.}
    \label{tab:uncertainty}
\end{table}

A summary of the uncertainty budget with its five uncorrelated contributions to the measurement
uncertainty of $p$ is shown in Table~\ref{tab:uncertainty} for the pCAVS \#2 data shown in Fig.~\ref{fig:example_decays}(b).
Error propagation of the measurement equation \eqref{eq:meas_equation} implies that the first three entries of Table~\ref{tab:uncertainty} depend on $p$.
In particular, the contribution in quadrature to $u(p)$
from $f_{\rm gl}$ is $|{\rm d}p/{\rm d}f_{\rm gl}|\times u(f_{\rm gl}) \approx p u(f_{\rm gl})$ because $f_{\rm gl} \ll 1$.
The last two entries in the table are independent of $p$, because $u(\Gamma)$ and $u(\Gamma_{\rm other})$ are independent of $p$, while $\Gamma \propto p$ and $\Gamma_{\rm other}=0$.
Combining all sources of uncertainty in quadrature, we find $p=41.5(1.2)$~nPa and $p=42.2(1.0)$~nPa for pCAVS \#1 and pCAVS \#2, respectively.
The relative uncertainties are 3.0~\% and 2.5~\%, respectively.
Moreover, the data in Table~\ref{tab:uncertainty} imply that for $p<40$ nPa the noise in $\Gamma-\Gamma_{\rm other}$ dominates in the uncertainty of $p$, while for $p>40$ nPa the uncertainty in loss rate coefficient $K_{{\rm H}_2}$ is the limiting factor in measuring $p$.
The uncertainty budget for pCAVS \#1 is the same as that for pCAVS \#2, except for a change in $u(\Gamma)$ to 0.78~nPa.

\begin{figure}
    \centering
    \includegraphics{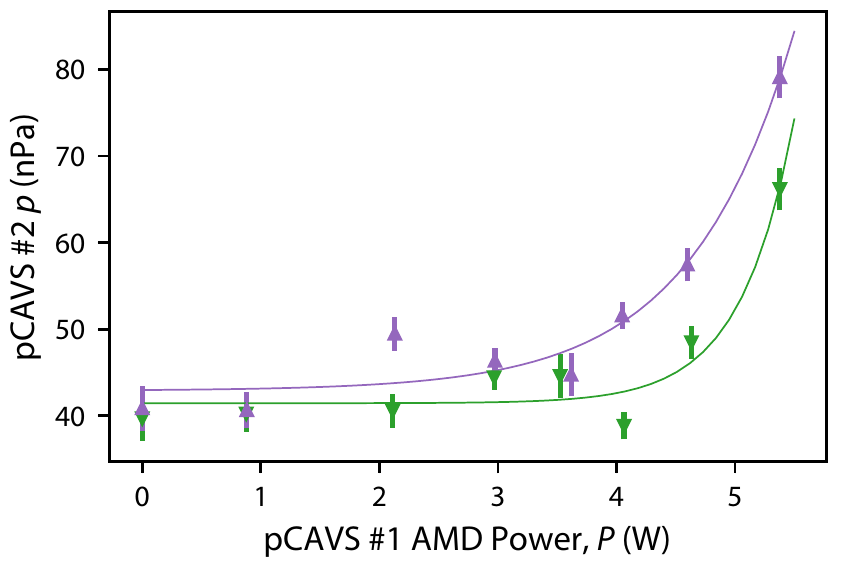}
    \caption{Measured pressures $p$ by pCAVS \#2 as functions of the power $P$ dissipated in the alkali-metal dispenser (AMD) of pCAVS \#1, with the latter's titanium shutter closed (green downward triangles) or open (purple upward triangles). Error bars are computed according to the uncertainty budget of Table~\ref{tab:uncertainty}. Curves are exponential functions fit to the data (see text).}
    \label{fig:outgassing}
\end{figure}

At base pressure near 40~nPa, we find that the alkali-metal dispensers (AMDs) affect the pressure in the vacuum chamber.
As described in Sec.~\ref{sec:description}, at the 400~$^\circ$C operating temperature of the AMD, we expect some outgassing from the dispenser.
Some fraction of this additional gas load is not pumped away by the NEG, and will escape through the differential pumping tube.
The additional gas load will scale exponentially with AMD temperature, which  in turn is determined by the balance of electrical power dissipated within the AMD ($\approx 3$~W) and the flow of heat out.
For our operating temperatures of about 400~$^\circ$C, radiative loss is negligible and the temperature rise of the AMD is proportional to the dissipated power.

We attempt to quantify the additional gas load by measuring the chamber pressure $p$ with pCAVS \#2 at constant AMD power while varying the AMD power $P$ of pCAVS \#1 to determine whether AMD outgassing affects the vacuum.
For this set of experiments, no $^7$Li atoms were loaded in pCAVS \#1.
Figure~\ref{fig:outgassing} shows the result with the pCAVS \#1 titanium shutter closed (green downward triangles), which lowers the conductance through the differential pumping tube, and open (purple upward triangles).
Both traces show a pressure increase that is exponential with dissipated power $P$, as expected from the observation that outgassing is exponential in temperature which, in turn, is proportional to $P$.
The pressure data is fit to $p = p_0+ q(P)/S$, where power-dependent (additional) outgassing rate $q(P) = q_0[\exp(P/P_0)-1]$ and $S= 10$~L/s, an estimate of the pumping speed to remove gas from pCAVS \#1.
The fit parameters $p_0$, $q_0$, and $P_0$ correspond to the chamber base pressure, the AMD outgassing rate coefficient, and the activation power for the outgassing process, respectively.

The data sets with the shutter closed and open in Fig.~\ref{fig:outgassing} were taken on different days.
On these days the base pressure of the vacuum chamber $p_0$ was slightly different due to changes in the laboratory environment.
At our typical operating power of 3~W, the resulting outgassing rates $q(P)$ are $2\times 10^{-8}$~Pa~L/s and $1\times10^{-9}$~Pa~L/s with the shutter open and closed, respectively.
The corresponding pressure increases are 2~nPa and 0.1~nPa, respectively.
These results suggest that the AMD is the predominant source of additional gas load; future versions of the pCAVS may incorporate a titanium AMD, which has a significantly lower outgassing rate compared to commercially available AMDs.\cite{Norrgard2018}
The 0.1~nPa increase when the shutter is closed suggests that the AMD is heating the surrounding vacuum components, causing them to  outgas, albeit at a considerably smaller rate.
Indeed, our thermometers observe an increase of  2~K for every 1~A of additional AMD current $i$.
We therefore conclude that to operate the pCAVS in the XHV regime ($<1$~nPa) both a better AMD and better thermal management are required.

Two limitations to operating the current pCAVSs in the XHV regime are due to the uncertainties in $\Gamma_{\rm Majorana}$ and $\Gamma$ (see Tab.~\ref{tab:uncertainty}).
Here, we have taken a conservative estimate of the uncertainty due to Majorana loss as further modeling is required.
Modeling would also determine its time dependence.
If this dependence is different from an exponential, it can be separated from the background-gas induced losses.\cite{Eckel2018}
If Majorana loss is both exponential in time and as large as  estimated in Tab.~\ref{tab:uncertainty}, future pCAVS designs could incorporate an Ioffe-Pritchard trap, which has no magnetic field zero and thus Majorana loss is suppressed.\cite{Bergeman1987,Bagnato1987}

The type-A statistical uncertainty of $\Gamma$ could be improved by more averaging, with the only relevant constraint being total integration time.
For example, the traces in Fig.~\ref{fig:example_decays}(b) required a collection time of roughly 3.5~h in order to achieve a relative uncertainty of $u(\Gamma)/\Gamma\approx 0.01$.
The required integration time scales with the inverse of pressure as $p\propto \Gamma$.
For a pressure of $0.4$~nPa, a hundred times better than our current base pressure, the integration time becomes 14~d to achieve the same $\approx 1\,\%$ relative uncertainty.
Improvements to the pCAVSs could be made by increasing the signal-to-noise ratio $N/u(N)$: we estimate that our current noise level is a factor of three higher than the atom-shot-noise limit.
As the integration time also scales as $N^{-1/2}$, loading up to $10^7$ atoms in the MOT, rather than the $10^5$ atoms with the current setup, will reduce the integration time by a factor $10$.
Such an improvement in atom number could be achieved by better aligning the laser, differential pumping tube, and diffraction grating chip.
Together, these improvements could reduce the  integration time at lower pressures.

\section{Conclusion}

We demonstrated the operation of two portable cold-atom vacuum sensors (pCAVS) based on  laser-cooled $^7$Li sensor atoms.
The sensors act as primary standards of pressure in the UHV domain.
We measure pressures as low as 40~nPa with 3~\% relative $k=1$ accuracy, and the two pCAVSs agreed in their measurement of this vacuum.
The largest contributor to this uncertainty is due to the uncertainty in the theoretical calculations of sensor-atom loss rate coefficient $K_{{\rm H}_2}$ of $^7$Li atoms colliding with H$_2$, the dominant contaminant molecule in a UHV vacuum.\cite{Makrides2019,Makrides2020,Makrides2022Ea}
Of the 40~nPa of pressure, approximately 1~nPa is caused by outgassing due to the operation of the pCAVSs.
We believe that this problem can be mitigated by switching to titanium AMDs and by employing more efficient thermal management.

The theory value for $K_{{\rm H}_2}$ has yet to be validated.
At NIST, efforts are underway to measure $K_{{\rm H}_2}$ with a laboratory-scale CAVS by creating a known pressure of order $10^{-7}$~Pa using a calibrated flowmeter and dynamic expansion vacuum metrology system.\cite{Scherschligt2017}
As part of this effort, we plan to attach one of our two pCAVSs to the dynamic expansion system, allowing us to compare four standards: the laboratory CAVS, the pCAVS, and flowmeter/dynamic expansion system.

Finally, we have shown that two or more pCAVSs can be used to locate leaks in a vacuum chamber.
Here, we found a leak with rate $10^{-6}$~Pa~L/s within one of the pCAVSs.
Multiple pCAVSs can triangulate a leak.
While we used only two pCAVSs operating from the same laser, our current setup allows for four.
Such leak triangulation by ionization gauges is limited by drifts in ion-gauge calibration; because the pCAVS is primary, no such calibration drifts exist.
With multiple pCAVSs having a 1~nPa to 2~nPa level of accuracy, impressively small leaks on the order of $10^{-9}$~Pa~L/s or $10^{-16}$~mol/s can be triangulated in large vacuum chambers, like beam lines.

\section*{Acknowledgments}
The authors thank P. Elgee and A. Sitaram for technical assistance with an early prototype and J. Klos, K. Rodriguez, and W. Tew for their comments on the manuscript.


%

\end{document}